\documentclass{WileyMSP-template}
\usepackage{multirow}
\usepackage{multicol}
\usepackage{caption}
\usepackage{subcaption}
\usepackage{xcolor}
\usepackage[normalem]{ulem}
\usepackage{bm}
\usepackage{bigints}

\begin{document}

\pagestyle{fancy}
\rhead{\includegraphics[width=2.5cm]{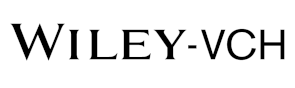}}

\title{Electronic structure of (organic-)inorganic metal halide perovskites: the dilemma of choosing the right functional}

\maketitle

% Author: Please give full first and last names for authors and include * after the name of all corresponding authors

\author{Cecilia Vona*}
\author{Dmitrii Nabok}
\author{Claudia Draxl*}

\dedication{}

% Affiliations: Please provide adacemic titles (Prof. or Dr.) for all authors where applicable, and include an institutional email address for all corresponding authors
\begin{affiliations}
C. Vona, Dr. D. Nabok, Prof. Dr. C. Draxl\\
 Institut f\"ur Physik and IRIS Adlershof, Humboldt-Universit\"at zu Berlin, Berlin, Germany\\
European Theoretical Spectroscopic Facility (ETSF) \\
Cecilia.Vona@physik.hu-berlin.de; Claudia.Draxl@physik.hu-berlin.de
\end{affiliations}

% Keywords: Please provide a minimum of three and a maximum of seven keywords, separated by commas


% Abstract should be written in the present tense and impersonal style (i.e., avoid we), and be at most 200 words long (For the moment around 180)

\begin{abstract}
Organic-inorganic metal halide perovskites (HaPs) are intensively studied for their light-harvesting properties. Owing to the interplay between strong electron-electron interaction and spin-orbit coupling (SOC), their quantitative theoretical description is still a challenge as evidenced by the wide variety of results available in literature. Here, various methodologies for computing their electronic structure are evaluated,
also accounting for SOC. More specific, the $GW$ approach as well as variants of the hybrid functionals PBE0 and HSE are at the center of our investigations. For both functionals, we explore methods to determine the mixing parameter $\alpha$, and for HSE, we investigate the impact of the screening-parameter $\omega$. An extensive investigation of PbI$_2$, a precursor of many HaPs, leads to the conclusion that hybrid functionals with $\alpha$ tuned by the density-based mixing method are most suitable for obtaining band gaps comparable to $G_0W_0$ results. Moreover, this methodology is transferable to CsPbI$_3$, and the same behavior is expected for the entire family of lead-iodine perovskites. 
\end{abstract}

% Text: Please use section headings and subheadings as specified below. For communications, all section headings apart from Experimental Section should be removed
% Please make the first reference to a display item bold: \textbf{Figure 1}
% Do not abbreviate Figure, Equation, etc.; display items are always singular, i.e., Figure 1 and 2.
% Equations are always singular, i.e., Equation 1 and 2, and should be inserted using the {equation} environment, not as graphics
% Please do not use footnotes in the text, additional information can be added to the Reference list.

\section{Introduction}
Organic-inorganic metal halide perovskites (HaPs) are in the focus of various optoelectronic applications, be it efficient solar-cells devices, lasers, or detectors. For instance, solar cells employing them as active layers, exhibit a power-conversion efficiency (PCE) of 25.5$\%$ \cite{Https://www.nrel.gov/pv/assets/images/efficiency-chart.png}, and perovskites/Si tandem cells have reached within less than five years a PCE of 29.5$\%$, a higher value than for any other solar cell on the market. 

Perovskites have the chemical formula ABX$_3$, where A is an organic or inorganic cation, like methylammonium (MA$^+$), formamidinium (FA$^+$) or Cs$^+$; B is a divalent metal cation, like Pb$^{2+}$ or Sn$^{2+}$; and X is an anion of the halogen group  (Cl$^-$, I$^-$,  Br$^-$). Among them, most suitable for solar-cell applications are lead-iodine based perovskites (APbI$_3$). Many current studies are focused on finding a replacement for the toxic element Pb, thereby preserving or even improving the physical properties. Double perovskites, combining a trivalent and a monovalent  metal cation, have been proposed as an alternative, also thanks to good stability and an energy gap that is highly tunable in the visible range \cite{Ghasemi2021Lead-freeApplications,Filip2016BandExperiment}. Other strategies to improve the material's stability consist in mixing organic and inorganic cations \cite{Pellet2014Mixed-organic-cationHarvesting,Lee2015FormamidiniumCell,Yi2016EntropicCells} or alternating layers of 3D and 2D HaPs \cite{Jung2019EfficientPoly3-hexylthiophene}.

Due to a growing complexity, resulting in large unit cells, one aims at accurate and, at the same time, computationally efficient approaches with high predictive power. The variety of methods to overcome the band-gap problem of semi-local density-functional theory (DFT), spans from employing hybrid functionals with varying amount of exact exchange to the $GW$ approximation of many-body perturbation theory (MBPT), or a combination of both. All of them are computationally involved. Another particular challenge is to capture the strong spin-orbit coupling (SOC) effects \cite{Even2013ImportanceApplications} that significantly lower the band gap. The goal of this work is to provide insight into the performance of different methods and to determine a fully \textit{ab initio} procedure to accurately compute the band gaps of lead-based perovskites and their precursor PbI$_2$. 

\begin{figure}
    \centering
    \includegraphics[width=10 cm]{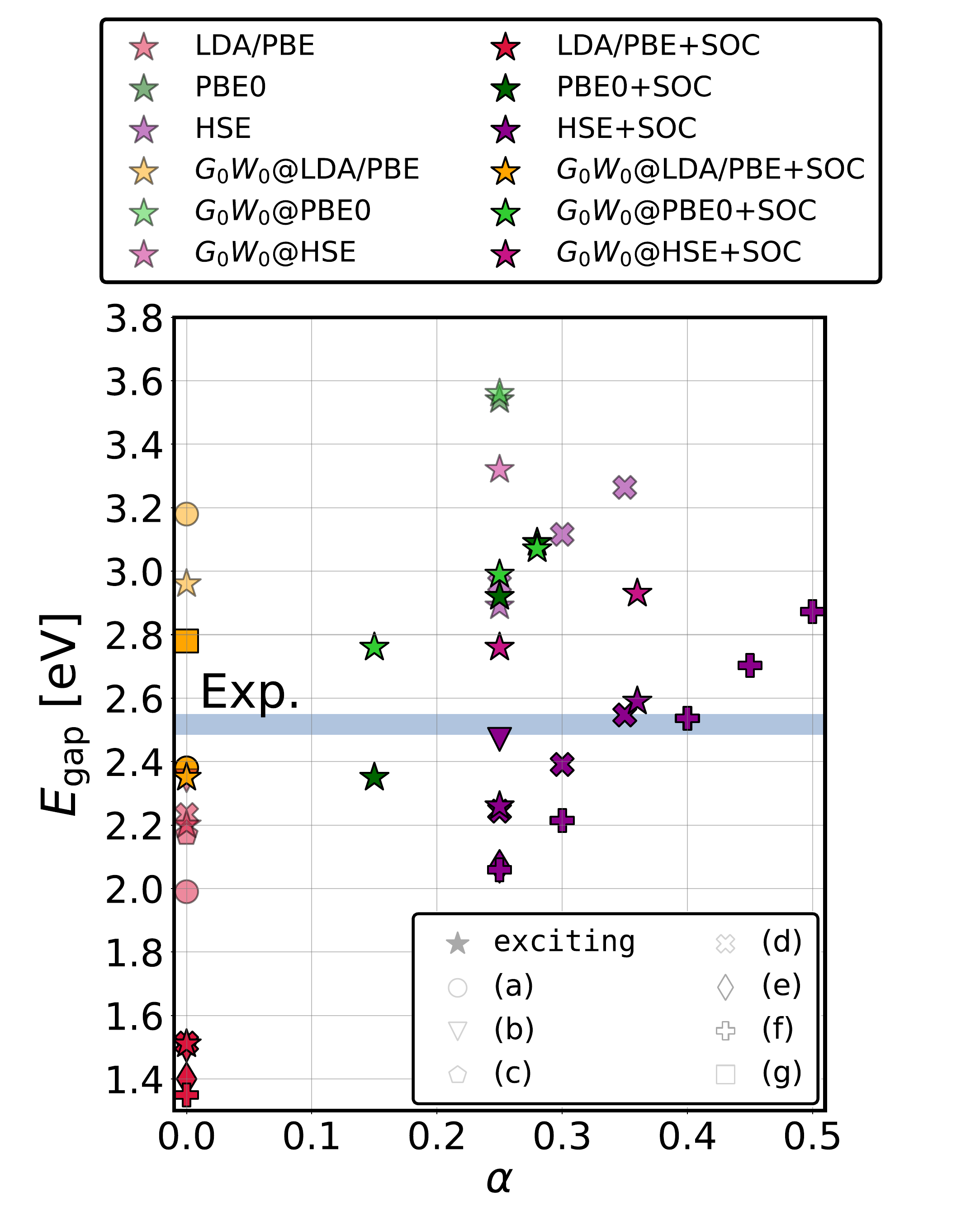}
        \caption{Energy gap of bulk PbI$_2$, computed with different methodologies and codes, as a function of the amount of Haretree-Fock exchange $\alpha$ in the hybrid functionals PBE0 and HSE. The range of experimental values (Ref. \cite{Gahailler+ExcitonicIodide,Ahuja2002ElectronicIodide}) is indicated by the blue bar. Results of this work (computed with \texttt{exciting}) are depicted by stars, the others are collected from literature, {\it i.e.}, references (a) \cite{Toulouse2015Frenkel-likeI2}, (b) \cite{Zhou2015SingleSpin-orbital-coupling}, (c) \cite{Du2019Perseverance7.69}, (d) \cite{Shen2018ElectronicStudy}, (e) \cite{Yagmurcukardes2018ElectronicMonolayer}, (f)  \cite{Borghesi2019ThePerovskites}, (g) \cite{Shen2019Excitonic}. These include computations with and without SOC, performed with semi-local and hybrids functionals, as well as $G_0W_0$ calculations using different functionals as starting point.}
        \label{Figure1} 
\end{figure}

To highlight the problem, we summarize in \textbf{Figure \ref{Figure1}} a collection of band-gap results for PbI$_2$ as computed with a variety of methods \cite{AlexanderL.FetterQuantumSystems,Mahan2000Many-ParticlePhysics}. Obviously, the enormous spread is far from being satisfactory, suggesting a lack of predictive power. We will get back to these results step by step in Section \ref{sec:results}. To shed light onto the role of different approaches, we are going to reexamine the performance of commonly used semi-local and hybrid functionals for the class of the perovskite materials. In this context, we mention other works to determine a reliable method to compute the electronic structure of HaPs. Among them, several focus on calculations based on MBPT \cite{Umari2014RelativisticApplications,Brivio2014RelativisticAbsorbers,Leppert2019TowardsGW,Wiktor2017PredictivePerovskites}, others employ HSE \cite{Feng2014CrystalHSE06, Kim2020AFunctional} or aim at improving hybrid functionals by tuning their parameters \cite{Menendez-Proupin2014Self-consistentPerovskite,Bischoff2019NonempiricalPerovskites}. Here, we place a focus on the dependency of the results on the functionals' parameters. In particular, we apply the \textit{dielectric dependent hybrid (DDH) method} \cite{Alkauskas2008BandCalculations} and the \textit{density-based mixing (DM) method} \cite{Marques2011Density-basedFunctionals}. We aim at finding optimal and transferable mixing parameters for two most popular hybrid functionals, {\it i.e.}, PBE0 and HSE. Moreover, we investigate the dependence of one-shot $GW$ ($G_0W_0$) on different parametrizations as the starting point. 
All the calculations are performed using the full-potential all-electron computer package \texttt{exciting} \cite{Gulans2014Exciting:Theory}. The computational details are provided in the Appendix.

\section{Methodology} 

\subsection{{Hybrid functionals and their parameters}}

To put our work in a bigger context, we shortly summarize the status of hybrid functionals. First suggested by Becke in 1993 \cite{Becke1993ATheories}, they combine a (semi-)local functional with a fraction of Hartree-Fock (HF) exchange. This way, they partially compensate for the missing exchange-correlation (xc) discontinuity and the self-interaction problem. Part of their success in reasonably predicting band gaps is associated with the error cancellation between the effects of the HF method, which tends to highly overestimate them, and KS DFT that often underestimates them \cite{BylanderGoodApproximation,Zhang2020HybridDevelopments}. After the \textit{half-and-half} hybrid functional \cite{Becke1993ATheories}, based on this idea, several such functionals were swiftly developed \cite{Becke1993Density-functionalExchange,Becke1996Density-functionalMixing,Perdew1996RationaleApproximations}. Among them, PBE0, proposed independently in two works \cite{Adamo1999TowardModel,Ernzerhof1999AssessmentFunctional}, is often termed the first fully \textit{ab initio} hybrid xc-functional. The PBE0 energy reads as follows:
\begin{equation}
    E_{\mathrm{xc}}^{\mathrm{PBE0}}=E_{\mathrm{xc}}^{\mathrm{PBE}}+\alpha(E_{\mathrm{x}}^{\mathrm{HF}}-E_{\mathrm{x}}^{\mathrm{PBE}}),
\end{equation}
where the mixing parameter $\alpha$ is set to 0.25. This fraction of HF exchange was justified by Perdew and coworkers \cite{Perdew1996RationaleApproximations} by applying G\"orling-Levy perturbation theory \cite{GorlingCorrelation-energyExpansion}.  Nevertheless, in Ref. \cite{Perdew1996RationaleApproximations}, the authors suggest to optimize the mixing parameter for each system and property. PBE0 is frequently used for solids, though generally overestimating band gaps of typical semiconductors \cite{Gerosa2018AccuracyExperiments,Borlido2019Large-ScaleSolids}. This overshoot is caused by the inclusion of the long-range (LR) tail of the Coulomb interaction that is expected to be effectively screened in periodic systems \cite{KohnDensityAtoms}.

To overcome this problem, Heyd, Scuseria, and Ernzerhof presented a screened hybrid functional \cite{Heyd2003HybridPotential}, known as HSE (first as HSE03), in which only the short-range (SR) part of the HF exchange is considered. The elimination of the LR part, besides improving the physical description of solid systems, is computationally convenient due to faster convergence with respect to the number of $\bf{k}$-points \cite{Schlipf2011HSEGdN}. 
To compute the SR HF exchange and substitute it by a fraction of SR PBE exchange, the Coulomb operator is split in a SR part and a LR part. To do so, one possibility is to make use of the error function and its complementary: 
\begin{equation}\label{Coul-split}
    v(r)=v^{\mathrm{SR}}(r)+v^{\mathrm{LR}}(r)=\frac{\mathrm{erfc}(\omega r)}{r}+\frac{\mathrm{erf(\omega r)}}{r}
\end{equation}

where $\omega$ is the screening parameter. This parameter assumes different values in the different forms of HSE, such as HSE03 \cite{Heyd2003HybridPotential,Heyd2006Erratum:8207} and HSE06 \cite{ Krukau2006InfluenceFunctionals}. As explained in the Ref. \cite{Heyd2006Erratum:8207}, HSE03 considers different values of the screening parameter for the HF-SR and the PBE-SR part ($\omega^{\mathrm{HF}}=0.15/\sqrt{2}\approx0.106\:\mathrm{a}_0^{-1}$ and $\omega^{\mathrm{PBE}}=0.15 \cdot 2^{1/3}\approx0.189\:\mathrm{a}_0^{-1}$), while HSE06 \cite{ Krukau2006InfluenceFunctionals},
\begin{equation}
     E_{\mathrm{xc}}^{\mathrm{HSE06}}=  E_{\mathrm{xc}}^{\mathrm{PBE}} + \alpha[E_{\mathrm{x}}^{\mathrm{HF,SR}}(\omega)-E_{\mathrm{x}}^{\mathrm{PBE,SR}}(\omega)],
\end{equation}
uses the same value for both ($\omega=0.11 \mathrm{a}_0^{-1}$) and $\alpha$=0.25 (same as in PBE0). HSE06 shows great performance for small- and medium-gap semiconductors, while underestimating the gaps of wide-gap materials \cite{Gerosa2018AccuracyExperiments,Borlido2019Large-ScaleSolids}. In this work, we focus on the most widely used hybrid functionals, PBE0 and HSE06 and evaluate two recently proposed first-principles techniques to optimize their parameters.

\subsubsection{Dielectric dependent hybrid functionals}\label{DDH} 
Marques and coworkers \cite{Marques2011Density-basedFunctionals} proposed to adopt the inverse of the static dielectric constant as the mixing parameter for PBE0, $\alpha=\varepsilon_{\infty}^{-1}$. The same idea is behind all the DDH methods developed later \cite{Shimazaki2008BandMethod, Shimazaki2014Dielectric-dependentCalculations,Koller2013HybridParameter,Skone2014Self-consistentSystems}. 
This relation is obtained by comparing the self-energy $\Sigma$ of MBPT with the generalized KS equation that is solved in case of hybrid functionals \cite{Seidl1996GeneralizedProblem, BylanderGoodApproximation}.
For the large set of materials, investigated in Ref. \cite{Marques2011Density-basedFunctionals}, the DDH method shows an improvement in the calculation of energy gaps with respect to PBE0, reducing the average error from 29.42$\%$ to 16.53$\%$ as compared to experiment. This average error, however, does not reflect that PBE0 performs considerably well for intermediate-sized band gaps, while it tends to overestimate (underestimate) the gaps of narrow (wide-gap) band-gap materials that are characterized by strong (weak) electronic screening \cite{Zhang2020HybridDevelopments}. The application of this method can be somewhat ambiguous, since there are several methods to compute the dielectric constant. A common approach to access the dielectric function is based on linear-response theory in the framework of time-dependent DFT \cite{Runge1984PHYSICALSystems}. In this approach, one has to rely on the choice of an exchange-correlation kernel to evaluate the response function. Thereby, the random-phase approximation (RPA) \cite{111CorrelationGas} has been shown to perform well \cite{Liu2020AssessingNiO}, due to error cancellation between the underestimation of the gap by, {\it e.g.} PBE, and the absence of electron-hole interactions \cite{Liu2020AssessingNiO}. Unfortunately, for complex materials, also RPA calculations can become expensive. The complexity in determining the dielectric constant motivated the authors of Ref.~\cite{Marques2011Density-basedFunctionals} to propose the DM method as an alternative.  

\subsubsection{Density-based mixing method}
\label{DM1} 

The idea of the DM method is to link the mixing parameter to a global estimator of the gap, obtained as the average of a local estimator over the unit cell. Local band-gap estimators, depending on $|\nabla n|/n$, have been proposed over the years in different contexts \cite{GutleCorrelationGap, Krukau2008HybridSeparation,Jaramillo2003LocalFunctionals}, and the idea of averaging it over the unit cell, to obtain a global estimator, was previously employed in the meta-GGA of Tran and Blaha (TBE09) \cite{Tran2009AccuratePotential}. The quantity proposed in Ref. \cite{Marques2011Density-basedFunctionals} has the following form:
\begin{equation}\label{gbar}
    \bar{g}=\frac{1}{V_{\mathrm{cell}}}\int_{V_{\mathrm{cell}}}\sqrt{\frac{|\nabla n(\boldsymbol{r})|}{n(\boldsymbol{r})}}\mathrm{d}\boldsymbol{r}.
\end{equation}
In PBE0, the mixing parameter $\alpha$ and $\bar{g}$ are connected by the following linear relation,
\begin{equation}
     \alpha_{\mathrm{PBE0-DM}}=-1.00778+1.10507\; \bar{g},
\end{equation}
whereas in the case of HSE, the proposed relation is in the fourth power of the estimator:
\begin{equation}
    \alpha_{\mathrm{HSE06-DM}}=0.121983+0.130711 \;\bar{g}^4.
\end{equation}

For the band gaps of of the materials set used in Ref. \cite{Marques2011Density-basedFunctionals}, average errors of 14.37$\%$ and 10.36$\%$ have been found for PBE0-DM and HSE06-DM, respectively, compared to 29.42$\%$ for PBE0 and 16.92$\%$ for HSE, thus showing a substantial improvement.  Another evaluation of the method for a large set of materials can be found in Refs. \cite{Borlido2019Large-ScaleSolids,Borlido2020Exchange-correlationLearning}.
A further advantage of this method, is that the estimator in Equation (\ref{gbar}) is easy to compute. As found in Ref.~\cite{Marques2011Density-basedFunctionals}, the functional used to compute the density has negligible influence on $\bar{g}$, thus this quantity can be calculated at the end of a ground-state calculation performed with a (semi-)local functional. The same method has been investigated also by D. Koller \textit{et al.} \cite{Koller2013HybridParameter}, who used a different definition of $\bar{g}$ and did not find any satisfactory relation between $\bar{g}$ and the mixing parameter $\alpha$. An advanced version of the method, suitable for interfaces, has been proposed more recently \cite{Borlido2018LocalInterfaces}.

\subsection{$G_0W_0$ approach}

As derived by L. Hedin~\cite{Hedin1965NewProblem}, the $GW$ approximation is a powerful method of MBPT with high predictive power for many classes of materials \cite{Martin2004ElectronicMethods}.
This approximation yields the quasiparticle (QP) energies that can be obtained in, {\it e.g.}, direct and inverse photoemission experiments. It is a common practice to formulate the $GW$ approximation in a perturbative way, based on the mean-field solutions provided by DFT. This scheme, commonly known as {\it single-shot $GW$} or {\it $G_0W_0$,} leads to the following expression for the QP energies~\cite{HybertsenElectronEnergies}
\begin{equation}
    \varepsilon_{n\boldsymbol{k}}^{\mathrm{QP}}= \varepsilon_{n\boldsymbol{k}}^0+Z_{n\boldsymbol{k}}\langle \varphi_{n\boldsymbol{k}}^0|\Sigma(\varepsilon_{n\boldsymbol{k}}^0)-v_{\mathrm{xc}}|\varphi_{n\boldsymbol{k}}^0\rangle,
    \label{eq:e_qp}
\end{equation}
in which $\Sigma$ is the electron self-energy that plays the role of a generalized (non-local, energy dependent, and non-Hermitian) exchange-correlation potential, and $Z_{n\mathbf{k}}$ is the QP renormalization factor:
\begin{equation}
Z_{n\boldsymbol{k}} = \Big[1-\frac{d}{d\omega}\langle\varphi_{n\boldsymbol{k}}^0|\Sigma(\omega)|\varphi_{n\boldsymbol{k}}^0\rangle_{\omega=\varepsilon_{n\boldsymbol{k}}^0}\Big]^{-1}. 
\end{equation}
In these equations, $\varphi_{n\boldsymbol{k}}^0$ and $\varepsilon_{n\boldsymbol{k}}^0$ are the eigenfunctions and eigenvalues, respectively, of the independent-particle problem solved by employing the xc functional $v_{\mathrm{xc}}$. The solution of Equation~(\ref{eq:e_qp}) implies a dependence of the QP energies on the choice of the underlying KS eigenstates. The latter must provide a reasonable approximation for the QP states such that a perturbative treatment is valid. Otherwise, the single-shot QP corrections alone will not be sufficient to improve band gaps, electronic binding energies, or even the order of bands. Hybrid functionals are typically good starting points for the solution of the QP equation. For semiconductors and insulator, HSE06 is often the favorite choice, while $G_0W_0$ on top of PBE0 tends to maintain the inherited overestimation of band gaps \cite{Fuchs2007QuasiparticleScheme}. The starting-point dependence can be remedied by adopting self-consistent $GW$ approaches \cite{VanSchilfgaarde2006QuasiparticleTheory,Grumet2018BeyondCalculations}. We do not, however, apply them here. 

\section{Systems under investigation}
\begin{figure}[h!]
    \centering
    \includegraphics[width=16 cm]{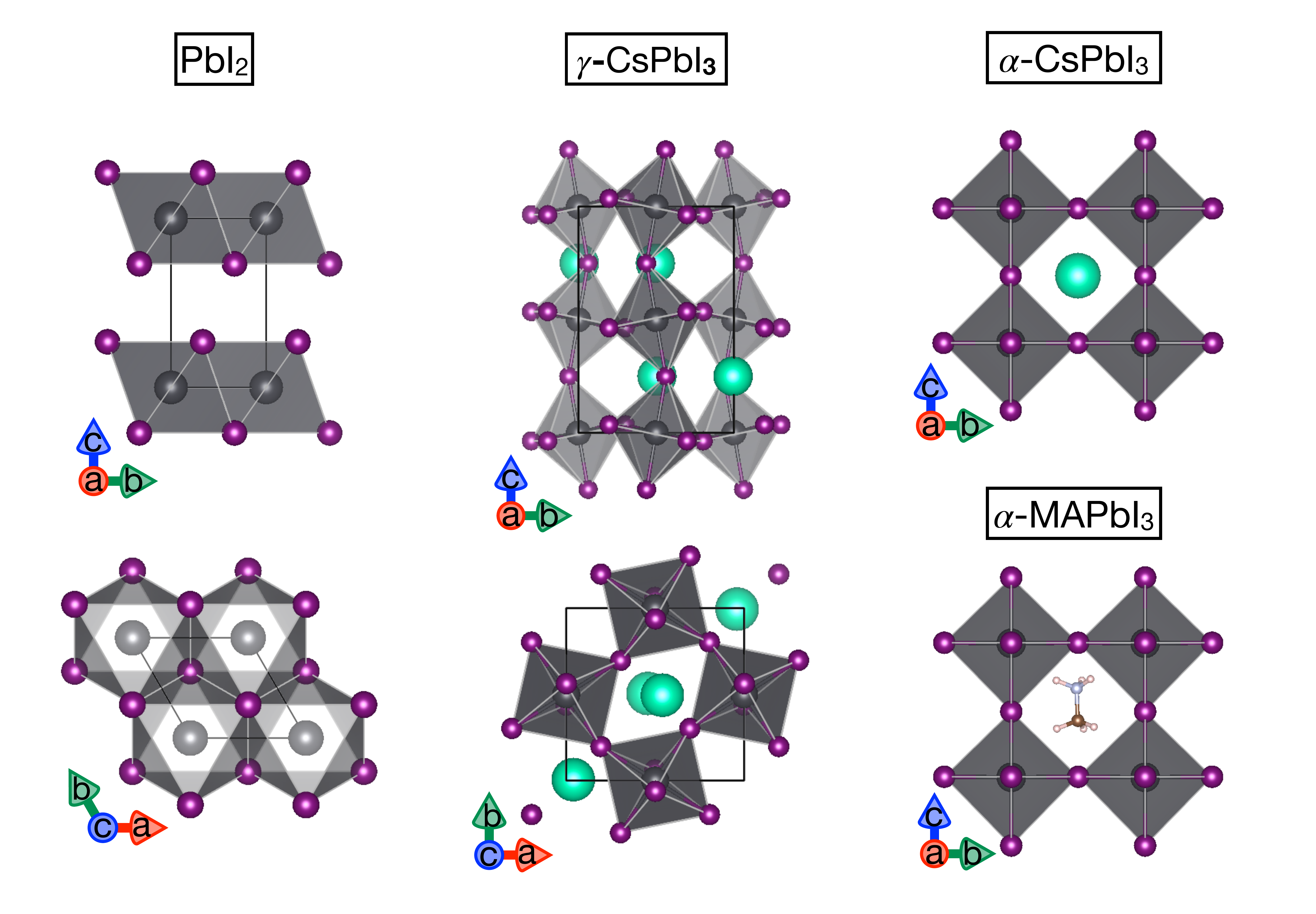}
        \caption{Unit cells of the four systems under investigations. To highlight differences between $PbI_{2}$ and $\gamma$-CsPbI$_3$, different orientations are shown in the respective top and bottom panels. } 
        \label{Figure2}
\end{figure}

We consider CsPbI$_3$ in its orthorhombic (space group Pnma) and cubic phase (space group Pm-3m), MAPbI$_{3}$ (space group Pm-3m), as well as bulk PbI$_2$ (space group P-3m1). Following the nomenclature of Ref. \cite{Stoumpos2013SemiconductingProperties}, we refer to the orthorhombic structure as $\gamma$-phase and to the cubic one as $\alpha$-phase. The unit cells are shown \textbf{Figure~\ref{Figure2}}. For $\gamma$-CsPbI$_3$, $\alpha$-CsPbI$_3$, and PbI$_2$ we adopt the experimental structural parameters \cite{Sutton2018CubicExperiment,Trots2008High-temperatureTriiodoplumbates,Paloszt1990TheTransition}. For $\alpha$-MAPbI$_3$, the hydrogen positions cannot be resolved by X-ray-based crystallographic methods; therefore, we adopt the structure from Ref. \cite{Vorwerk2018Exciton-DominatedPerovskites}. The higher symmetry of the $\alpha$-phase (high-temperature phase) of HaPs is related to a dynamical disorder in the octahedral tilts. For a good theoretical description, without explicitly accounting for dynamical effects, it is convenient to consider lower-symmetry unit cells such as the $\gamma$-phase structure \cite{Even2015Solid-stateSemiconductors, Quarti2016StructuralCells,Sutton2018CubicExperiment}. Therefore, we will compare the results obtained for $\gamma$-CsPbI$_3$ to experimental data, while we consider $\alpha$-CsPbI$_3$ and $\alpha$-MAPbI$_3$ to investigate the effects of the crystalline structure and the type of cation. 

\section{Results and discussion}\label{sec:results}

\subsection{Electronic structure of APbI$_3$ and PbI$_2$}

\subsubsection{Semi-local DFT and SOC}\label{electronic}
The four systems under investigation are made of lead and iodine, heavy atoms that are responsible for significant SOC effects. In Table \ref{tabbandgap}, the fundamental band gaps computed with PBE and PBE+SOC are shown. The SOC-induced band-gap reduction is dramatic in all systems, ranging from 0.69 eV in PbI$_2$ to 1.13 eV in $\alpha$-CsPbI$_3$. Comparing the values for $\alpha$-CsPbI$_3$ and  $\gamma$-CsPbI$_3$, the sensitivity of the band gap to the crystal symmetry and details of the atomic arrangement in the PbI$_3$ cage are apparent. Contrarily, the nature of the cation (organic or inorganic) plays a less decisive role in the electronic band structure in the surrounding of the Fermi level. The differences between the band gaps of $\alpha$-CsPbI$_3$ and $\alpha$-MAPbI$_3$, caused by the deviation of $\sim$0.01 \AA{} in the Pb-I bond length, are found to be within 100 meV, for both PBE and PBE+SOC.

\begin{figure}[h!]
    \centering
    \includegraphics[width=18 cm]{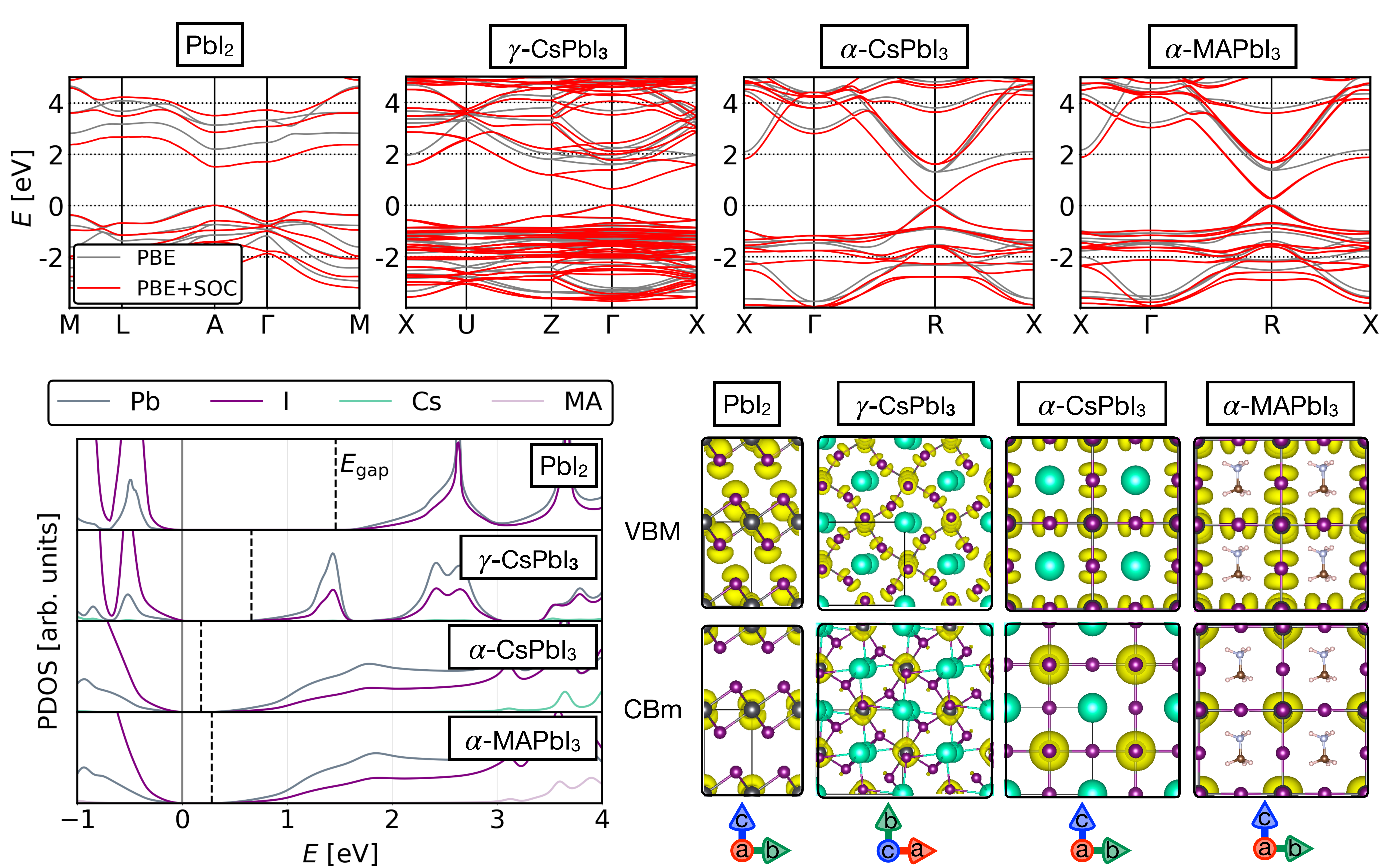}
        \caption{Electronic properties of PbI$_2$, $\gamma$-CsPbI$_3$, $\alpha$-CsPbI$_3$, and $\alpha$-MAPbI$_3$. Top: Band structures computed with PBE (gray) and PBE+SOC (red). In each panel, the bands are aligned at the VBM that is set to 0. Bottom left: Atom-resolved partial DOS computed with PBE+SOC. In the case of $\alpha$-MAPbI$_3$, the DOS of the MA molecule is considered for comparison with Cs in the other structures. Bottom right: Square modulus of the KS wavefunctions at VBM and CBm, computed with PBE+SOC.}
        \label{Figure3}
\end{figure}

For further analysis, we present in the top row of \textbf{Figure~\ref{Figure3}} the band structures along selected high-symmetry paths. 
The bands computed with PBE (gray lines) and PBE+SOC (red lines) are aligned at the valence-band maximum (VBM) to facilitate comparison. In the perovskites structure, SOC effects in the valence bands are minimal, whereas in PbI$_2$, especially at $\Gamma$, they are responsible for the splitting of several bands. Importantly, the reduction of the gap here is a consequence of the lifting of the degeneracy at the conduction-band minimum (CBm), while in PbI$_2$ the splitting occurs in the second unoccupied band. Also in this case, the difference between the band structures of the two $\alpha$-phase perovskites is minimal, in contrast to that of the $\gamma$-phase. The energy gaps of the $\alpha$-phase perovskites are located at R, in the $\gamma$-phase at $\Gamma$. In the $\alpha$-phase perovskites, SOC lifts the triple degeneracy of the CBm, by splitting it into a single state (CBm) and a doubly degenerate state. In the $\gamma$-phase, SOC is not only responsible for splitting the doubly-degenerate CBm but also for changing the order of the states at $\Gamma$.

Being found critically important, from this point on, we proceed with the discussion of the electronic properties, always taking SOC into account (unless noted otherwise). For a deeper analysis, we consider the partial density of states (bottom left of Figure~\ref{Figure3}). We observe (i) that in all four materials, the VBM (CBm) is dominated by iodine (lead) $p$-orbitals, and (ii) the organic/inorganic cation states do not contribute to the band-gap region. Concerning (i), in all considered materials, the VBM also exhibits a contribution from Pb-$s$ orbitals as evident from Figure S1 of the Supporting Information where the orbital-resolved DOS are presented. The orbital character is also clear from the KS wave functions presented in the bottom right of Figure~\ref{Figure3}. Notably, the CBm orbitals do not reveal the typical $p$-orbital shape but rather the relativistic $p^{1/2}$ and $p^{3/2}$ forms~\cite{WhitePICTORIALATOMS}, reflecting the strong SOC. The latter and the composition of the band-edge states, which PbI$_2$ shares with APbI$_3$, provide already justification for taking PbI$_2$ as a representative compound for exploring the electronic structure with different methodologies. Finally, we will demonstrate that all conclusions drawn for precursor PbI$_2$ can be transferred to the lead-iodide perovskites.

\begin{table}
    \caption{Electronic band gaps (in eV) of PbI$_2$, $\gamma$-CsPbI$_3$, $\alpha$-CsPbI$_3$, and $\alpha$-MAPbI$_3$, calculated with and without SOC and employing different methods.}
    \label{tabbandgap}
    \begin{tabular}[htbp]{@{}lcccc@{}}
        \hline
        & \rm{PbI$_2$} &\rm{$\gamma$-CsPbI$_3$} & \rm{$\alpha$-CsPbI$_3$}& \rm{$\alpha$-MAPbI$_3$}\\
        \hline
        Exp.& 2.55$^{a)}$, 2.485$^{b)}$& 1.73$^{c)}$,1.67$^{d)}$ &- &1.69$^{e)}$\\ 
        PBE     & 2.20 &1.58& 1.31 & 1.35 \\
        PBE+SOC & 1.51 & 0.63 & 0.18 & 0.28\\
        PBE0    & 3.54 & 2.75 & 2.32 &-\\
        PBE0+SOC& 2.92 &1.86  & 1.27 &-\\
        HSE06   & 2.89 & 2.13 & 1.75& -\\
        HSE06+SOC & 2.26 & 1.25 & 0.70& -\\
        $G_0W_0$@PBE & 2.96 & 2.17  & 1.98 & -\\
        $G_0W_0$@PBE+SOC & 2.35 & 1.32 & 0.94 & -\\
        $G_0W_0$@PBE0 & 3.56 & 2.84 & - & -\\
        $G_0W_0$@PBE0+SOC & 2.99 & 1.99 & - & -\\
        $G_0W_0$@HSE06 & 3.32 & 2.54 & - & -\\
        $G_0W_0$@HSE06+SOC& 2.76 &1.72 & - & -\\
        \hline
    \end{tabular}
    \begin{flushleft}
        $^{a)}$ Ref.~\cite{Gahailler+ExcitonicIodide} \quad 
        $^{b)}$ Ref.~\cite{Ahuja2002ElectronicIodide} \quad
        $^{c)}$ Refs.~\cite{McmeekinACells, Eperon2014FormamidiniumCells} \quad
        $^{d)}$ Ref.~\cite{Singh2019InvestigationFilm}\quad
        $^{e)}$ Ref.~\cite{Quarti2016StructuralCells}
    \end{flushleft}
\end{table}

\subsubsection{Hybrid functionals and one-shot $GW$}\label{standardmethods}

In this section, we discuss the performance of standard methods (either DFT or MBPT) in predicting the electronic band gaps for the studied materials. In Table \ref{tabbandgap}, the corresponding values computed with PBE, PBE0, and HSE are shown together with those obtained by $G_0W_0$ on top of them, both with and without SOC. The amount by which SOC reduce the gap, is comparable for all methods. For $\alpha$-MAPbI$_3$, no results for hybrid functionals and $GW$ are provided since the organic/inorganic cation has no direct influence on the band gap in 3D lead-iodide perovskites. For the high-temperature phase of CsPbI$_3$ ($\alpha$-phase), no experimental reference is available, however, as for the other APbI$_3$ materials, we can expect the gap to be close to the one of $\gamma$-CsPbI$_3$ \cite{Quarti2016StructuralCells}. All the theoretical results summarized in Table~\ref{tabbandgap}, turn out to be significantly lower than the experimental value of $\gamma$-CsPbI$_3$. As shown for MAPbI$_3$ \cite{Quarti2016StructuralCells}, the reason is that the local octahedral environment is not symmetric and shows structural similarity to the $\gamma$-phase. This is why we will not use the $\alpha$-phase to evaluate the different methods.

By comparing the calculated gap (including SOC) of $\gamma$-CsPbI$_3$ with the experimental references (Table~\ref{tabbandgap}), we observe that $G_0W_0$@HSE, giving a value of 1.72 eV, reproduces experiment best. Moreover, PBE0 overestimates the gap by only 130 meV ($\sim7\%$). HSE and $G_0W_0$@PBE lead to an underestimation by 420 meV ($\sim25\%$) and 350 meV ($\sim21\%$), respectively. For PbI$_2$, the best agreement with experiment is achieved by $G_0W_0$@PBE, showing a slight overestimation by 135 meV ($\sim5\%$). Also HSE and $G_0W_0$@HSE perform well with values being by 225 meV ($\sim9\%$) too low and by 210 meV ($\sim8\%$) too high, respectively. PBE0 overestimates the gap by 370 meV ($\sim14\%$). For both materials, $G_0W_0$@PBE0 overshoots slightly more than PBE0 (by 70 meV in PbI$_2$ and 130 meV in $\gamma$-CsPbI$_3$). 

\begin{figure}
    \centering
    \includegraphics[width=18 cm]{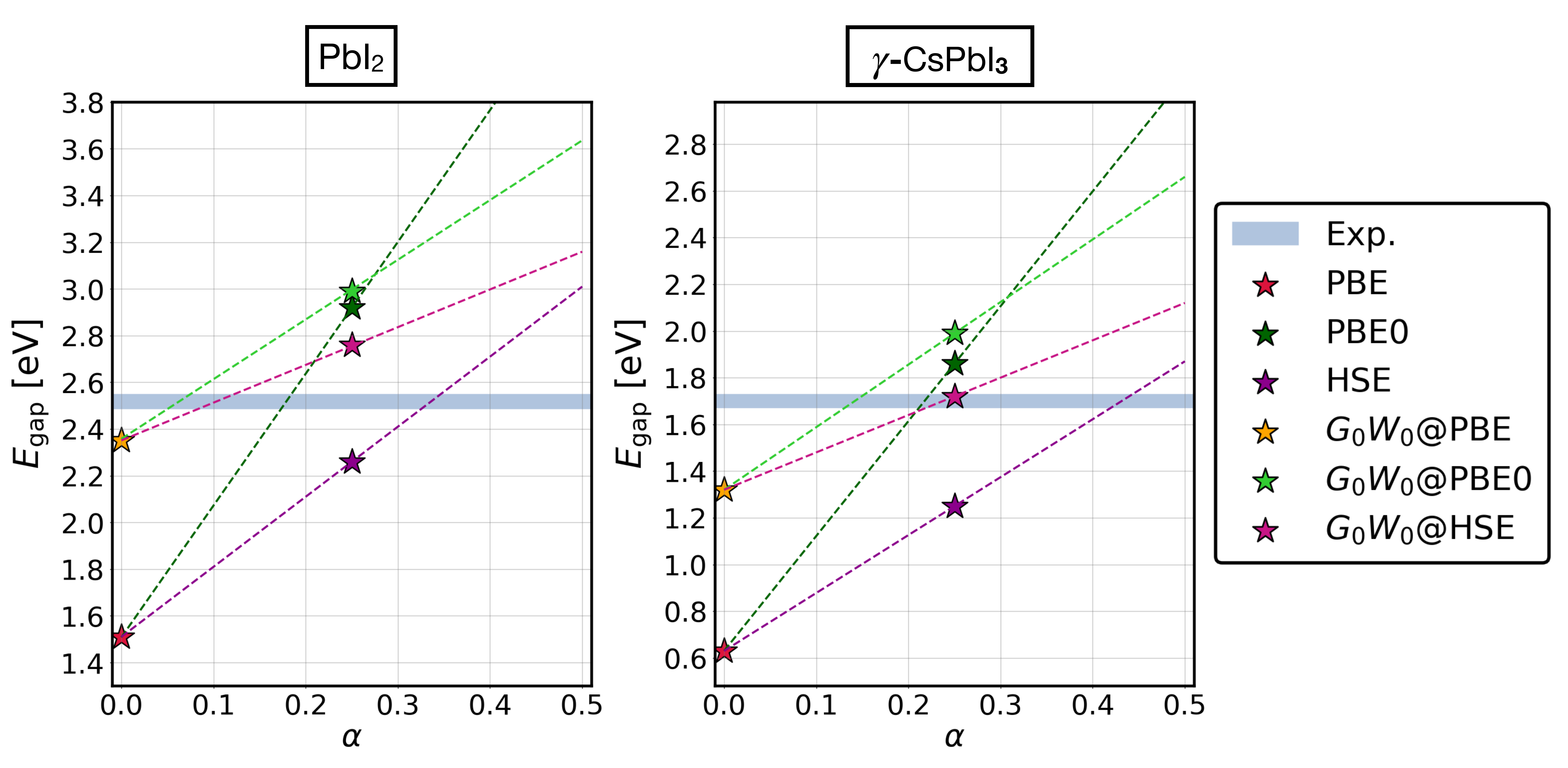}
        \caption{Energy gaps (considering SOC) of PbI$_2$ (left) and $\gamma$-CsPbI$_3$ (right) as a function of mixing parameter $\alpha$ (see also Table~\ref{tabbandgap}). The dashed lines are guides to the eye that reflect the known linear behavior. }
    \label{Figure4}
\end{figure}

In \textbf{Figure~\ref{Figure4}}, the energy gaps from Table~\ref{tabbandgap} (considering SOC) are plotted. The dashed lines indicate how the results obtained with the hybrid functionals and $G_0W_0$ on top of them change with respect to the mixing parameter $\alpha$. It should be noted that comparison between experimental and computed values bear some uncertainties due to, {\it e.g.}, crystal structure, presence of defects or thermal effects. Moreover, in our calculations, electron-phonon coupling is not taken into account \cite{Wright2016Electron-phononPerovskites}. Therefore, we also focus on comparing different computational methods. From Figure~\ref{Figure4}, some trends for the two materials can be observed: (i) The ascending order of the gap computed with the different methods is the same. (ii) The differences between the methods are comparable. These observations justify transferability of the methods within materials of this class.

Summarizing, $G_0W_0$@HSE appears overall to perform best for predicting the band gap of the materials under investigation. However, it is computationally very demanding, which leads us to explore methods to tune the parameters of the hybrid functionals PBE0 and HSE.

\subsection{Mixing parameters for PBE0 and HSE}
\subsubsection{Dielectric dependent hybrid method}
The key quantity of the DDH method is the electronic dielectric constant $\varepsilon_{\infty}$. As discussed in Section \ref{DDH}, RPA@PBE gives a good estimate. As SOC effects are decisive for describing the electronic properties in HaPs and their precursors, it is important to include SOC also when computing the dielectric response. Such
calculations can be computationally expensive for complex materials such as perovskites. Therefore, we mimic the effects of SOC through a scissors operator, taken as the difference between the energy gaps obtained with PBE+SOC and PBE. From the band structures in the top panel of Figure~\ref{Figure3}, we can immediately observe that the scissors approximation is fully valid for PbI$_2$, since the lowest unoccupied band is rigidly shifted, while the same is not true in the case of HaPs. To nevertheless justify and validate the use of the scissors shift within the DDH method, we compare the energy gaps and the mixing parameters obtained when employing PBE+SOC and PBE+scissors, respectively, for the calculation of the dielectric constant within the independent particle (IP) approximation. The results are shown in Table \ref{table2}.  As expected, for PbI$_2$, they differ by a small amount only, leading to the same values for the optimized mixing parameter and for the energy gap. In the HaPs, instead, the usage of a scissors shift leads to a bigger dielectric constant. The differences in the mixing parameter $\alpha$ and thus in the energy-gaps are, however, small for all the HaPs, {\it i.e.},  $\alpha$ being 0.01 bigger when SOC is accounted for explicitly. The energy gaps of $\gamma$-CsPbI$_3$ and $\alpha$-CsPbI$_3$ obtained by the two methods are within 0.05 eV. From this analysis we conclude that, for determining the mixing parameter and further the energy gap, the scissors operator is legitimate. Moreover, through this analysis we confirm that SOC is fundamental for the DDH method, since by adopting PBE only, the final gap is up to 200 meV (for $\gamma$-CsPbI$_3$) bigger than the one obtained from PBE+SOC.

For the reasons just pointed out above, for tuning the PBE0 mixing parameter with the DDH method (PBE0-DDH), we compute the dielectric constant employing RPA@PBE, and we make use of a scissors correction to account for SOC effects. For PbI$_2$, the energy gap obtained with PBE0-DDH is equal to the one computed with $G_{0}W_0$@PBE, which underestimates the experimental gap by $\sim$5$\%$. Moreover, PBE0-DDH improves over PBE0 that shows an overestimation by $\sim$14$\%$ compared to experiment. Also for $\gamma$-CsPbI$_{3}$, the result of PBE0-DDH is comparable to that of $G_{0}W_0$@PBE (difference within 50 meV); however PBE0 can already reproduce the experimental gap well with a difference of only $\sim$7$\%$, considerably better than PBE0-DDH (difference $\sim$18$\%$).

From the results in Table~\ref{table2}, we observe that the organic/inorganic cation has a smaller effect on the mixing parameter than the crystal structure. This is in accordance with the electronic structures of the different compounds, analyzed in Section~\ref{electronic}. The effect of the structure is, however, moderate with  $\alpha$ ranging from 0.12 for $\alpha$-CsPbI$_3$ to 0.15 for PbI$_2$ and $\gamma$-CsPbI$_{3}$. By transferring the mixing parameter obtained for PbI$_2$ to compute the gap of $\alpha$-CsPbI$_3$ the error is 130 meV only (Table~\ref{DDH}), and for $\alpha$-MAPbI$_3$ we expect it to be even smaller since the $\alpha$ values are even closer.

\begin{table}
    \caption{ Dielectric constant obtained from different approximations, corresponding mixing parameter, $\alpha$, obtained by the DDH method, and energy gaps from PBE0 and $GW$@PBE0, employing the respective $\alpha$. The values marked by $^{\perp}$ are computed with the procedure shown in the Appendix, all other values result from the linear fits shown in Figure S2 in the  Supporting Information.}
    \label{table2}
    \begin{tabular}[htbp]{@{}llcccc@{}}
        \hline
        {\rm Material}&{\rm Method}  & {\rm $\varepsilon_{\infty}$} & {\rm $\alpha$} & {\rm PBE0($\alpha$)}& {\rm $G_{0}W_{0}$@PBE0($\alpha$)}\\
        \hline
        PbI$_2$&IP@PBE  & 7.31 & 0.14 & 2.30 & 2.71\\
        {} & IP@PBE+SOC & 8.27 & 0.12 & 2.18 & 2.66\\
        {} & IP@PBE+scissor & 8.46 & 0.12 & 2.18 & 2.66\\
        {} & RPA@PBE & 5.74 & 0.17 & 2.47 & 2.79\\
        {} & RPA@PBE+scissor & 6.88 & 0.15 & 2.35$^{\perp}$ & 2.76$^{\perp}$\\
        $\gamma$-CsPbI$_3$&IP@PBE & 6.02 & 0.17 & 1.47 & 1.82\\
        {} & IP@PBE+SOC & 7.10 & 0.14 & 1.32 & 1.73 \\
        {} & IP@PBE+scissor & 7.69 & 0.13 & 1.27 &1.70 \\
        {} & RPA@PBE & 5.13 & 0.19 & 1.56 & 1.88 \\
        {} & RPA@PBE+scissor & 6.46 & 0.15 & 1.37 & 1.76\\
        $\alpha$-CsPbI$_3$& IP@PBE & {6.43} & 0.15 & 0.83 & - \\
        {} & IP@PBE+SOC & 9.07 & 0.11 & 0.66 & - \\
        {} & IP@PBE+scissor & 9.77 & 0.10 & 0.62 & - \\
        {} & RPA@PBE & 5.57 & 0.18 & 0.96 & -\\
        {} & RPA@PBE+scissor & 8.38 & 0.12 & 0.70 & -\\
        $\alpha$-MAPbI$_3$& IP@PBE& {6.35} & 0.16 &- &- \\
        {} & IP@PBE+SOC & 8.23 & 0.12 & - & - \\
        {} & IP@PBE+scissor & 9.10 & 0.11 & - & - \\
        {} & RPA@PBE & 5.48 & 0.18 & - & - \\
        {} & RPA@PBE+scissor & 7.78 & 0.13 & - & -\\
        \hline
    \end{tabular}
\end{table}

\subsubsection{Density-based mixing method}
To the best of our knowledge, the DM method was not applied to HaPs before; neither were SOC effects taken into account. As we see from Table \ref{DM}, the effect of SOC on $\bar{g}$ is negligible. We attribute this finding to the fact that $\bar{g}$ is obtained by an average over the unit cell (Equation~(\ref{gbar})), as well as to the minor differences between the total electron densities of PBE and PBE+SOC, as SOC mainly affects the conduction bands. 

The DM method applied to PBE0 (PBE0-DM) leads to a larger overestimation of the gaps with respect to PBE0. In contrast, the DM method in combination with HSE (HSE-DM) improves over HSE for the gaps of both materials. For PbI$_2$, the overestimation is only 40 meV ($\sim$1.6$\%$), making this method the best choice overall. For $\gamma$-CsPbI$_3$, HSE-DM underestimates the gap by 130 meV ($\sim$8$\%$) which is comparable with PBE0 (difference of 130 meV). 

From Table~\ref{DM}, we observe that the parameters obtained for the four materials are in a small range (from 0.28 to 0.30 for PBE0-DM and from 0.36 to 0.38 for HSE-DM), a finding that, again would justify the use of the parameters found for PbI$_2$ for the other compounds. Moreover, the organic/inorganic cation has more influence than the crystalline structure, since the difference in the parameters between $\alpha$-CsPbI$_3$ and $\alpha$-MAPbI$_3$ is bigger than between $\alpha$-CsPbI$_3$ and $\gamma$-CsPbI$_3$. As  this difference is, however small, we again conclude that the organic cation does not have a major impact on the total electron density and its gradient.

\begin{table}
    \caption{Mixing parameters ($\alpha_{\mathrm{PBE0-DM}}$  and $\alpha_{\mathrm{HSE-DM}}$) computed with the DM method for PBE0-DM and HSE-DM and corresponding energy gaps in the DFT and MBPT framework. The values marked by $^{\perp}$ are computed with the procedure shown in the Appendix, all other values result from the linear fits shown in Figure S2 in the  Supporting Information. }
    \label{DM}
    \begin{tabular}[htbp]{@{}llccccccc@{}}
        \hline
        {\rm Material}&{\rm E$_{\mathrm{xc}}$} & {\rm $\bar{g}$ [a$_0^{1/2}$]} & {\rm $\alpha_{\rm{PBE0-DM}}$} & {\rm PBE0-DM} & {\rm $G_{0}W_{0}$@PBE0-DM}& {\rm $\alpha_{\rm{HSE-DM}}$} & {\rm HSE-DM} & {\rm $G_{0}W_{0}$@HSE-DM}\\
        \hline
        PbI$_2$ & PBE  & 1.166 & \multirow{2}{*}{0.28} & \multirow{2}{*}{3.09$^{\perp}$} & \multirow{2}{*}{3.07$^{\perp}$} & \multirow{2}{*}{ 0.36} &  \multirow{2}{*}{2.59$^{\perp}$}&  \multirow{2}{*}{2.93$^{\perp}$} \\
        {} & PBE+SOC &  " & & & & & &  \\  
        $\gamma$-CsPbI$_3$ & PBE & 1.176 & \multirow{2}{*}{0.29}  & \multirow{2}{*}{2.05} & \multirow{2}{*}{2.09} & \multirow{2}{*}{ 0.37} &  \multirow{2}{*}{1.54} &  \multirow{2}{*}{1.91} \\
        {} & PBE+SOC & " & & & & & &  \\ 
        $\alpha$-CsPbI$_3$ & PBE &1.168 & \multirow{2}{*}{0.28}  & \multirow{2}{*}{1.40} & \multirow{2}{*}{-} & \multirow{2}{*}{ 0.36} &  \multirow{2}{*}{0.93} &  \multirow{2}{*}{-} \\
        {} & PBE+SOC & 1.167 & & & & & &  \\
        $\alpha$-MAPbI$_3$ & PBE &1.187 & \multirow{2}{*}{0.30}  & \multirow{2}{*}{-} & \multirow{2}{*}{-} & \multirow{2}{*}{ 0.38} &  \multirow{2}{*}{-} &  \multirow{2}{*}{-} \\
        {} & PBE+SOC & 1.186 & & & & & &  \\  
        \hline
    \end{tabular}
\end{table}

\subsection{Screening parameter $\omega$ and $G_0W_0$ calculations}\label{screening} 
\begin{figure}[htbp]
    \centering
    \includegraphics[width=16 cm]{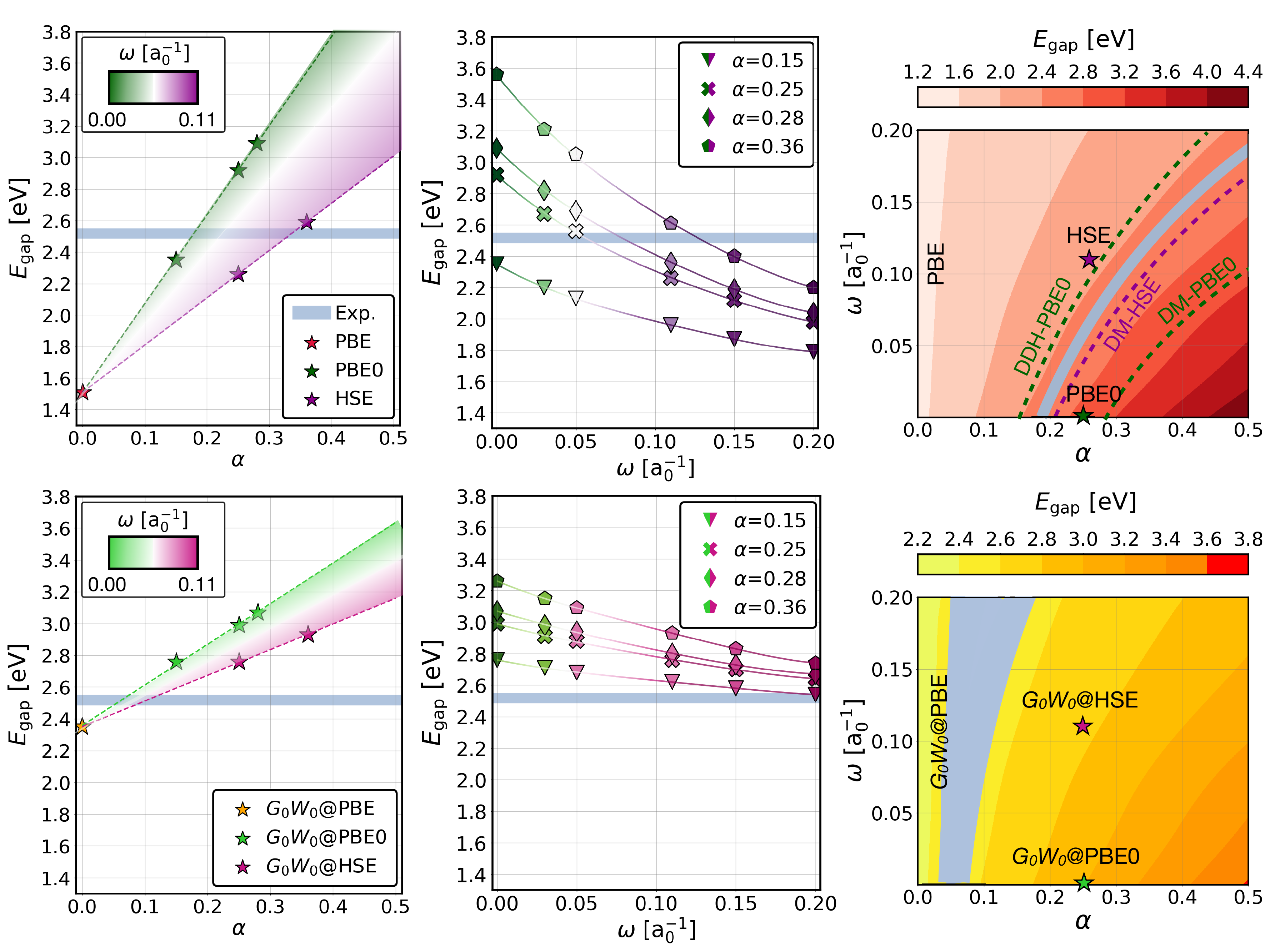}
        \caption{ Top: Band gap of bulk PbI$_2$ (considering SOC) as a function of $\alpha$ (left) and $\omega$ for different $\alpha$ (middle panel); and color map visualizing the effect of both (right). The bottom panels show the corresponding results when applying $G_0W_0$ on top. The color code inside the areas between the PBE0 ($G_0W_0$@PBE0) and HSE ($G_0W_0$@HSE) in the left panels is chosen to indicate the values of $\omega$. The same color coding is used in the middle panels. The here plotted values are summarized in the Supporting Information.}
        \label{Figure5}
\end{figure}

In this section, we investigate the impact of $\omega$, the second parameter of HSE, on the band gap of PbI$_2$. Moreover, we discuss the starting-point dependence of the $G_0W_0$ gaps and how they are influenced by the choice of the parameters $\alpha$ and $\omega$. To recall, in the limit $\alpha=0$, both PBE0 and HSE are identical to PBE. Moreover, by construction, HSE satisfies two limits: (i) For $\omega$=0, it coincides with PBE0, and (ii) for $\omega \rightarrow \infty$ it becomes equivalent to PBE as follows from Equation~(\ref{Coul-split}). The left-top panel of \textbf{Figure~\ref{Figure5}} visualizes the linear dependence of the PBE0 and HSE band gaps as a function of $\alpha$ (the symbols indicate the values from Tables \ref{tabbandgap}, \ref{table2}, and \ref{DM}). The color scheme shows how the gaps change with respect to $\omega$ for different values of $\alpha$. The latter can be seen more explicitly in the middle-top panel, in which for specific values of $\alpha$ (0.25 and the ones obtained from the tuning methods), the results for different values of $\omega$ are plotted. The $\alpha$-dependence gets less steep for bigger values of $\omega$, which reflects the definition of HSE, since for $\omega\rightarrow\infty$ all curves go asymptotically to the PBE value. These results are combined in the right-top panel of Figure~\ref{Figure5}, in which the color map encodes the energy gaps as a function of $\alpha$ and $\omega$. The isolines represent those combinations that reproduce the experimental, DDH, and DM band gaps, respectively (see also Tables~\ref{tabbandgap}, \ref{table2}, and \ref{DM}).
The DDH method provides a mixing parameter only for PBE0. Through the PBE0-DDH isoline we extract the $\alpha$ value for HSE ($\omega=0.11\; \mathrm{a}_0^{-1}$) which is $\alpha=0.28$.

We now perform the same analysis for $G_0W_0$. The results are included in Tables~\ref{table2} and \ref{DM} and, together with the values from Table~\ref{tabbandgap}, are plotted in
the bottom of Figure~\ref{Figure5}. 
%Again, the color map indicates the dependence of the gaps on $\omega$ and $\alpha$. 
Also in this case, the gaps change linearly with respect to $\alpha$, but in a much narrower range. Likewise, the middle panel shows the dependence of the results on $\omega$ for selected values of $\alpha$; while the color map on the right highlights the combined dependence on both parameters. From this analysis, one can deduce that the dependence of the energy gaps on $\alpha$ and $\omega$ in $G_0W_0$ is similar to that at the DFT level, but restricted to a smaller region. While on the DFT side, the gap changes in a range between 1.2 to 4.4 eV, on the $G_0W_0$ side, only between 2.2 to 3.8 eV. Overall, the starting-point dependence with respect to the hybrid parametrization is significant, even if the range is smaller. 
   
%%%%%%%%%%%%%%%%%%%%%%%%%%%%%%%%%%%%%%%%%%%%%%%
\section{Conclusions}
We have systematically investigated the performances of several methods of DFT and MBPT  to compute the band gaps of PbI$_2$ and APbI$_3$. To avoid mere benchmarking against experimental results, which comes with uncertainties, we also consider comparison between different theoretical approaches. We have verified that SOC is fundamental, irrespective of the method. This makes the calculations numerically challenging.
Overall, $G_0W_0$@HSE performs overall best but is computationally remarkably expensive. Among the approaches that involve the tuning of parameters, PBE0-DDH gives results comparable to $G_0W_0$@PBE, but is also cpu-intensive, partly owing to SOC. Instead, in the DM methods, SOC has no major effect on the estimator $\bar{g}$. Based on the PBE electron density to compute $\bar{g}$, makes this method accessible for complex materials such as APbI$_3$. At the DFT level, HSE-DM gives overall the best estimation of the energy gap, with a performance comparable with $G_0W_0$@HSE. 

Both tuning methods lead to similar mixing parameters for the four investigated materials, with the biggest difference found for PBE0-DDH, where $\alpha$=0.12 for $\alpha$-CsPbI$_3$ and $\alpha$=0.15 for $\gamma$-CsPbI$_3$ and PbI$_2$. This means that the atomic species Pb and I have larger effects on the screening parameter and on the density gradient ($|\nabla n|/n$) than the crystalline structure and the (in)organic cation. This reflects the fact that the band gap region is dominated by $p$-Pb and $p$-I states, which also explains the transferability of the method from PbI$_2$ to APbI$_3$. Investigating the dependency of the PbI$_2$ energy gap on different combinations of $\alpha$ and $\omega$, we find that both parameters have significant impact in both hybrid functionals. This also applies when $G_0W_0$ is computed on top, however, less pronounced. Our results suggest that one can transfer our findings from the studied precursors to the respective HaPs compounds.

\section*{Appendix A: Computational details}\label{compdetail}

All the DFT and MBPT calculations are performed using the full-potential all-electron computer package \texttt{exciting} \cite{Gulans2014Exciting:Theory}. The code employs the linearized augmented planewave plus local orbitals ((L)APW+lo) basis \cite{LumePHYSICALTheory,UseCopper,SjostedtAnMethod,SinghGround-stateStates} to expand the Kohn-Sham wavefunctions. In the (L)APW+lo method, the unit cell is partitioned into two regions: muffin-tin (MT) spheres around the atomic nuclei of radius $R_{\mathrm{MT}}$, in which the basis functions are atomic-like functions, and the interstitial region between the MT spheres, in which the basis functions are planewaves. The MT radii are chosen to be $R_{\mathrm{Pb}}=2.9\ \mathrm{a}_0$, $R_{\mathrm{I}}=2.9\ \mathrm{a}_0$, $R_{\mathrm{Cs}}=2.9 \ \mathrm{a}_0$, $R_{\mathrm{C}}=1.1 \ \mathrm{a}_0$, $R_{\mathrm{N}}=1.0 \ \mathrm{a}_0$, and $R_{\mathrm{H}}=0.9 \ \mathrm{a}_0$. The large sphere sizes of Pb and I avoid any core leakage, which is highly important when SOC is taken into account. The basis-set size is determined by the number of planewaves used.
As the planewave cutoff, $G_{\mathrm{max}}$, depends on the muffin-tin radii, it is common practice to express the cutoff parameter as the dimensionless product $R_{\mathrm{MT}} \cdot G_{\mathrm{max}}$, where $R_{\mathrm{MT}}$ is the radius of the smallest sphere. In the DFT calculations (with PBE and hybrid functionals), $\boldsymbol{k}$ mesh and $R_{\mathrm{MT}}G_{\mathrm{max}}$ are chosen such to guarantee a numerical precision of the band gap within 20 meV for PbI$_2$, $\alpha$-CsPbI$_3$, and $\alpha$-MAPbI$_3$ and within 50 meV for $\gamma$-CsPbI$_3$. All $G_0W_0$ band gaps are converged up to 100 meV. In Table~\ref{parameters1}, the values of $G_{\mathrm{max}}$ and $R_{\mathrm{MT}}G_{\mathrm{max}}$ used for the investigated systems are summarized.

The electronic structure is computed employing the xc-functionals PBE, PBE0, and HSE as well as $G_0W_0$ on top of them. The $\mathbf{k}$-grids are shown in Table~\ref{parameters2}. The convergence of the energy gaps with PBE0 turned out slower than with PBE, HSE, and $G_0W_0$. Nevertheless, for PbI$_2$, the same grid is used in HSE and PBE0 calculations such to investigate the asymptotic behavior of HSE $\rightarrow$ PBE0 for $\omega$ $\rightarrow$ 0.
SOC is included if specified. In case of PBE, it is treated via the second variational scheme  \cite{Singh2005PlanewavesEdition}, where in the corresponding term of the Hamiltonian, $H_{\mathrm{SO}} \propto \partial v_0(r)/r\partial r \; (\boldsymbol{\sigma} \cdot \boldsymbol{L})$,  $v_0$ is the spherical component of the effective potential.
For the considered set of materials, the SOC treatment within the second-variation scheme requires almost all KS functions available to achieve precise results. The corresponding number of the empty states is significantly higher than that required to converge the hybrid  and $G_0W_0$ calculations.
Therefore, to maintain the same level of precision with and without SOC, we use in all these calculations the higher number of empty states, {\it i.e.}, those required for SOC, which are 300 for PbI$_2$, 1000 for $\gamma$-CsPbI$_3$ , 460 $\alpha$-CsPbI$_3$, and 1000 for $\alpha$-MAPbI$_3$. 

\begin{table}
\centering
    \caption{$G_{\mathrm{max}}$ and  $R_{\mathrm{MT}}G_{\mathrm{max}}$ values used for the different systems under investigation. Since $R_{\mathrm{MT}}G_{\mathrm{max}}$ depends on the sphere size, individual values are provided. The large parameter for Pb and I in $\alpha$-MAPbI$_3$ is owing to the big differences in sphere sizes between the heavy atoms and H, as explained in the text. }
    \label{parameters1}
    \begin{tabular}[htbp]{@{}lccccccc@{}}
        \hline
        & \multirow{2}{*}{$G_{\mathrm{max}}$ [a$_0^{-1}$]}&\multicolumn{6}{c}{$R_{\mathrm{MT}}G_{\mathrm{max}}$}\\
        & &{\rm Pb}&{\rm I }  &  {\rm Cs }& {\rm C} & {\rm N }& {\rm H }\\
        \hline
        PbI$_2$ &2.76 & 8 & 8 & - & - & -& - \\
        $\gamma$-CsPbI$_3$ & 2.07& 6 & 6 & 6 & -& - & -  \\
        $\alpha$-CsPbI$_3$ &2.41& 7 & 7 & 7 & -& - & - \\
        $\alpha$-MAPbI$_3$ & 3.55& 10.3 &10.3 & - & 3.9 & 3.5 & 3.2 \\ 
        \hline
    \end{tabular}
\end{table}

\begin{table}
\centering
    \caption{$\boldsymbol{k}$-mesh used for the calculation of the electronic structure and the dielectric constant (with and without SOC). In the $G_0W_0$ calculations, always the same grid as in the underlying DFT calculation is employed.}
    \label{parameters2}
    \begin{tabular}[htbp]{@{}lcccc@{}}
        \hline
        &PBE&PBE0& HSE & $\varepsilon_{\infty}$ (SOC)\\
        \hline
        PbI$_2$ &3x3x2 & 6x6x4 & 6x6x4 & 6x6x4 (6x6x4)\\
        $\gamma$-CsPbI$_3$ &2x2x1& 3x3x2 & 2x2x1 & 6x6x4 (6x6x4)\\
        $\alpha$-CsPbI$_3$ & 4x4x4 & 6x6x6 & 4x4x4  & 10x10x10 (16x16x16)\\
        $\alpha$-MAPbI$_3$ & 4x4x4 & - & -& 10x10x10 (16x16x16) \\ 
        \hline
    \end{tabular}
\end{table}

Calculations with hybrid functionals, consist of a nested loop \cite{Betzinger2010HybridPBE0,Schlipf2011HSEGdN}. In the outer loop, the non-local exchange is computed by employing a mixed-product basis \cite{AryasetiawanProduct-basisMatrices,Jiang2013FHI-gap:Method}; in the inner loop, the generalized KS matrix equation is self-consistently solved by updating in each step only the local part of the effective potential. SOC is included self-consistently in the inner loop, again through second variation. Since the gradient of a non local potential is not trivial to compute, the effective potential in $H_{\mathrm{SO}}$ employs PBE \cite{Wang2017AnCoupling}. This is justified since this contribution is small. 
 
 Regarding HSE, the SR part of the HF exchange is computed by the difference of the total HF exchange and the HF-LR contribution as done by Schlipf and coworkers \cite{Schlipf2011HSEGdN}. The authors show that by doing so, in the limit of $\boldsymbol{q}+\boldsymbol{G}\rightarrow 0$, the singularities that would occur in the full HF exchange and HF-LR cancel out by Taylor expanding the exponent coming from the LR contribution, thus leading to the constant $\pi/\omega^{2}$. This approximation was also used by us in Ref. \cite{Aggoune2021FingerprintsExperiment}.  However, this expression implies numerical difficulties to study the asymptotic behavior of HSE for $\omega \rightarrow0$ caused by the necessity to employ extremely dense $\boldsymbol{k}$-point grids. To overcome this problem, we have derived an alternative expression to compute the SR part of the bare Coulomb potential in the limit of  $\boldsymbol{q}+\boldsymbol{G}\rightarrow 0$. More details are given in Appendix B.
 
 In our $G_0W_0$ calculations \cite{Nabok2016AccurateMethod}, SOC effects are incorporated by applying the second-variation procedure after the QP corrections to the Kohn-Sham energies have been computed. This is an approximate but computationally efficient way for treating SOC in $GW$. As discussed in Ref. \cite{Aguilera2013Spin-orbitInsulators}, this scheme may lead to wrong predictions in materials where SOC induces band-inversion such as topological insulators. In other systems like the ones investigated here, it is expected to produce results consistent with those from a more rigorous treatment.

Calculations of the dielectric functions are performed using both the independent-particle approximation and the RPA kernel \cite{Rubio2009Time-dependentTheory}. To evaluate the dielectric constants, the parameters are chosen such that the mixing parameter $\alpha$ of PBE0 is determined with a precision of $5 \cdot 10^{-3}$. The $\boldsymbol{k}$ meshes used to compute $\varepsilon_{\infty}$ are shown in Table~\ref{parameters2}. In the RPA calculations, the parameter {\it gqmax} that governs local-field effects is chosen 2 for all systems. The number of empty states is 10 for PbI$_2$, 20 for $\alpha$-CsPbI$_3$, and $\alpha$-MAPbI$_3$ and 100 for $\gamma$-CsPbI$_3$. These values are reduced with respect to those used for the electronic structure since SOC is not taken into account. The parameter $\bar{g}$ of the DM method (Equation~(\ref{gbar})) is computed from the PBE density. We verified that, using the parameters of Tables~\ref{parameters1} and \ref{parameters2}, $\alpha$ is determined with a precision of $5 \cdot 10^{-3}$. % lead to the same value for $\alpha$ (differences below $5 \cdot 10^{-3}$).} {\red Do you mean this?}

The crystal structures displayed in Figure \ref{Figure2} and the KS wavefunctions in Figure \ref{Figure3} have been produced with the software package VESTA \cite{Momma2011Data}.

\section*{Appendix B: Treatment of the singularity in the Coulomb potential} \label{AppendixA}

The Fourier transform of the short-range part of the Coulomb potential employed in the HSE  functional is given by
\begin{equation}
    v_{\mathrm{C}}(\boldsymbol{q+G})=\frac{4\pi}{|\boldsymbol{q+G}|^2}\bigg(1-\mathrm{e}^{-\frac{|\boldsymbol{q+G}|^2}{4\omega^2}}\bigg).
\end{equation}
It is easy to show \cite{Schlipf2011HSEGdN} that 
\begin{equation}
    \lim_{\boldsymbol{q+G}\rightarrow 0} v_{\mathrm{C}}(\boldsymbol{q+G})  = \pi/\omega^2.
\end{equation}
However, one quickly faces numerical instabilities when studying the parametric dependence of HSE results for small values of $\omega$. Therefore, the case of $\boldsymbol{q+G}\rightarrow 0$ and $\omega \rightarrow 0$ requires a special treatment. 

In this work, we estimate the limit by isotropic averaging in a small region around the $\Gamma$ point of the Brillouin zone (BZ). Our goal is to compute the integral
\begin{equation} \label{eq:integral_1}
    I_{\mathrm{s}}=\frac{1}{V_{k}}\bigintsss_{V_k}{\mathrm{d}\boldsymbol{q}}\,v_{\mathrm{C}}(\boldsymbol{q}),
\end{equation}
where $V_k = \Omega_{BZ}/N_k$ is a small volume surrounding $\Gamma$, $\Omega_{BZ}$ is the volume of the BZ, and $N_k$ is the total number of the $\mathbf{k}$-points used to sample the BZ. To be able to compute the integral analytically, we replace $V_k$ with a sphere of radius
\begin{equation}
    R_k=\bigg(\frac{3\Omega_{\mathrm{BZ}}}{4\pi N_k}\bigg)^3.
\end{equation}
Using the notation $\boldsymbol{q} \equiv \boldsymbol{q+G}$, $q \equiv |\boldsymbol{q+G}|$, and $\beta=1/4\omega^2$, Equation~(\ref{eq:integral_1}) turns into
\begin{equation} \label{eq:integral_2}
    I_{\mathrm{s}} = \frac{1}{V_{k}}\bigintsss_0^{R_k}dq \, 4\pi q^2 \; \frac{4\pi}{q^2}\bigg(1-\mathrm{e}^{-\beta q^2}\bigg) = \frac{16\pi^2}{V_{k}}\bigg[R_{k}-\sqrt{\frac{\pi}{4\beta}}\mathrm{erf}(\sqrt{\beta}R_{k})\bigg].
\end{equation}
It can be shown, that for big values of $\omega$ (small values of $\beta$), the integral correctly recovers the value $\pi/\omega^2$ from Ref.~\cite{Schlipf2011HSEGdN}. Our approach allows us to study the parametric dependence of the HSE results on both $\alpha$ and $\omega$ in the entire parameter space.

\medskip
\textbf{Supporting Information} \par %Please delete the Suppporting Information statement if it is not applicable. Please supply Supporting Information in another file. Supporting information should not be provided in .tex format
Supporting Information is available from the Wiley Online Library or from the author.

%%%%%%%%%%%%%%%%%%%%%%%%%%%%%%%%%%%%%%%%%%%%%%%
% Acknowledgements
\medskip
\textbf{Acknowledgements}
Work supported by the German Research Foundation within the priority program SPP2196 {\it Perovskite Semiconductors}, project Nr. 424709454 and the European Community’s Horizon 2020 research and innovation program under the Marie Skłodowska-Curie Grant Agreement No. 675867. We acknowledge the North-German Supercomputing Alliance (HLRN) for providing computational resources and thank Ronaldo Rodrigues Pela and Wahib Aggoune for critical reading of the manuscript. C. V. thanks Andris Gulans, Christian Vorwerk and Daniel Speckhard for fruitful discussions.

%%%%%%%%%%%%%%%%%%%%%%%%%%%%%%%%%%%%%%%%%%%%%%%
\medskip
\textbf{Conflict of Interest}
The authors declare no conflict of interest.

\medskip
\textbf{Open Research}
Input and output files are openly available in the NOMAD Repository \cite{NOMAD2019} at the following link https://dx.doi.org/10.17172/NOMAD/2021.10.26-1.

%%%%%%%%%%%%%%%%%%%%%%%%%%%%%%%%%%%%%%%%%%%%%%%
% References
% Use the following code if you wish to generate your bibliography with BibTeX;
% replace the string "MSP-template" below with the name(s) of
% the BibTeX data base(s) you want to use.
% The resulting bibliography-output (the content of the .bbl file)
% must be pasted back into this file before submission.
% Please also include your BibTeX data base file(s) in your submission
% so that we can re-run BibTeX if necessary.
%
\bibliographystyle{MSP}
%\bibliography{references.bib}

% Figures/tables and captions
% Permission statements are required for all figures reproduced or adapted from previously published articles/sources. Please also ensure that all necessary permissions to reproduce images have been received
% Please remove these statements for original figures

%%%%%%%%%%%%%%%%%%%%%%%%%%%%%%%%%%%%%%%%%%%%%%%
% Table of contents entry should be 50 - 60 words long
% Image should be 55 mm broad and 50 mm high or 110 mm broad and 20 mm high
\newpage
\begin{figure}
\textbf{Table of Contents}\\
\medskip
  \includegraphics[width=55mm ,height=50mm]{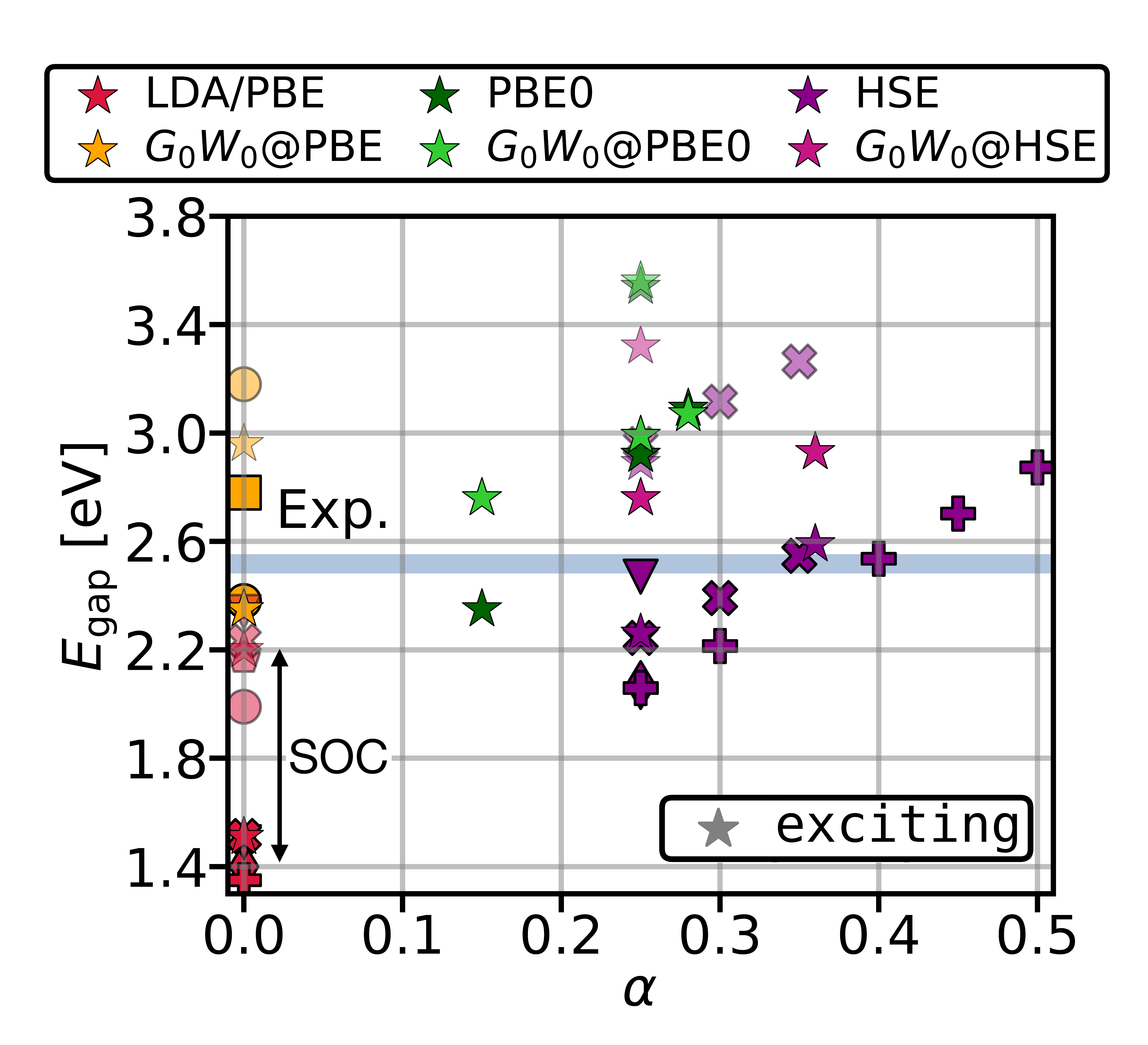}
  \medskip
  \caption*{Computing the electronic structure of organic-inorganic halide perovskites accurately, remains challenging. The spread of values that different available methodologies obtain for the energy gap, is enormous. This study shows, that properly parametrized hybrid functionals are suitable for this task, and that the procedure adopted for PbI$_2$, the precursor of lead-iodide perovskites, is transferable to the whole family of these materials. } \end{figure}

\end{document}